\documentclass[10pt,twocolumn,letterpaper]{article}

\usepackage{cvpr}              % CVPR template package
\usepackage{graphicx}          % For figures
\usepackage{amsmath}           % Math symbols
\usepackage{amssymb}           % Additional symbols
\usepackage{booktabs}          % For professional tables
\usepackage{multirow}          % For multi-row tables
\usepackage{caption}           % Custom captions
\usepackage{hyperref}          % Hyperlinks
\hypersetup{colorlinks=true, linkcolor=blue, citecolor=blue, filecolor=blue, urlcolor=blue}

\title{Guiding Treatment Strategies: The Role of Adjuvant Anti-Her2 Neu Therapy and Skin/Nipple Involvement in Local Recurrence-Free Survival in Breast Cancer Patients}

\author{Joe Omatoi \thanks{joe.omatoi@mercurial-ai.com} \quad 
Abdul M. Mohammed \thanks{abdul.mohammed@mercurial-ai.com} \quad 
Dennis Trujillo\thanks{Corresponding author: dennis.trujillo@mercurial-ai.com}\\
\\ 
Mercurial AI Inc
}
\begin{document}

\maketitle

%%%%%%%%% ABSTRACT
\begin{abstract}
This study explores how causal inference models, specifically the Linear Non-Gaussian Acyclic Model (LiNGAM) \cite{shhk06,sishkwhb11}, can extract causal relationships between demographic factors, treatments, conditions, and outcomes from observational patient data, enabling insights beyond correlation. Unlike traditional randomized controlled trials (RCTs), which establish causal relationships within narrowly defined populations, our method leverages broader observational data, improving generalizability \cite{mlkt22,bsvmdzj20}. Using over 40 features in the Duke MRI Breast Cancer dataset \cite{shgkgwm18}, we found that Adjuvant Anti-Her2 Neu Therapy increased local recurrence-free survival by 169 days, while Skin/Nipple involvement reduced it by 351 days. These findings highlight the therapy's importance for Her2-positive patients and the need for targeted interventions for high-risk cases \cite{dmtbh15}, informing personalized treatment strategies.
\end{abstract}

%%%%%%%%% BODY TEXT
\section{Introduction}
Breast cancer remains one of the most prevalent and challenging cancers worldwide, with an estimated 2.3 million new cases diagnosed globally in 2022 \cite{blsfssj24}. This figure represents 11.6\% of all cancer cases, making breast cancer the most commonly diagnosed cancer worldwide. The complexity of breast cancer lies in its heterogeneity, with various subtypes characterized by different molecular profiles, prognoses, and treatment responses \cite{hpco19}. This diversity emphasizes the need for personalized treatment approaches based on individual patient characteristics and tumor properties.

Given such complexities, the landscape of breast cancer research has been rapidly evolving, with increasing emphasis on personalized medicine and extraction of meaningful insights from large-scale observational data. While Randomized Controlled Trials (RCTs) have long been the gold standard for establishing causal relationships in medical interventions, there remains a significant gap in leveraging the wealth of information contained in observational datasets, such as the Duke MRI dataset, to extract causal insights that can guide clinical decisions \cite{ozlo23}.

Recent studies demonstrated substantial progress in understanding the complex interplay between interventions, patient characteristics, and outcomes in breast cancer. For instance, a group of researchers examined the impact of interventions like beta-blockers and angiotensin-converting enzyme inhibitors (ACEIs) on cardiovascular mortality risk for breast cancer patients, revealing that these treatments can be effective in reducing the incidence of conditions such as atrial fibrillation (AF). These studies also highlight the importance of demographic factors, such as age and race, in moderating treatment effects. For example, older Black female patients with severe breast cancer were found to be at a higher risk for developing AF, showing that intervention effectiveness varies across different populations \cite{gfdwlaa21}.  

Similarly, research on breast cancer treatments, such as chemotherapy and neoadjuvant systemic therapy (NST), has demonstrated that tumor characteristics (e.g., size, stage, and lymph node involvement) and patient demographics (e.g., age) can significantly influence treatment outcomes \cite{zzljdtwhcyla24}. These examples underscore the progress made in understanding how interventions interact with various patient characteristics, but they also point to a critical gap of untapped potential in observational datasets.  

The potential of causal inference methods to address key healthcare concerns has been increasingly recognized in recent years, particularly in overcoming the limitations of RCTs, which are often constrained by narrow entry conditions that limit generalizability. Being able to extract causal relationships from observational data helps us overcome these narrow entry conditions by enabling the study of a broader population and even allowing the possibility of combining multiple datasets within a single study. This approach also helps us explore more features, since not all factors can be included in an RCT or some set up may be too expensive to perform \cite{bsvmdzj20}. The examples above underline the importance of exploring treatment-demographic interactions to optimize patient outcomes, yet, despite the availability of rich datasets like Duke MRI, there has been no attempt to systematically extract causal effects from these observational datasets. This points to a substantial opportunity for future research to leverage causal inference methods and better understand these complex dynamics \cite{glwrwsaabbdhjlmmmrorsswwak21, sn22, ozlo23}.

This study utilizes the demographic, clinical, pathology, genomic, treatment, outcomes, and other data portion of the Duke MRI Breast Cancer dataset. "The dataset is a single-institutional, retrospective collection of 922 biopsy-confirmed invasive breast cancer patients, over a decade" \cite{shgkgwm18}. 

Through exploring the untapped potential of observational data, specifically within the Duke MRI dataset, we uncovered new and highly impactful findings: Adjuvant Anti-Her2 Neu Therapy was found to increase the number of days to the last local recurrence-free assessment by 169 days, on average. In contrast, Skin/Nipple involvement decreased this duration by 351 days, on average.

\section{Methodology}

\subsection{Dataset and Preprocessing}
The Duke MRI Breast Cancer dataset is a retrospective collection of 922 patients with invasive breast cancer. Key preprocessing steps included:
\begin{itemize}
    \item {Dropping irrelevant and redundant features}
    \item {Imputing missing values with mean/mode}
    \item {Encoding categorical variables using one-hot and ordinal encoding.}
\end{itemize}

Redundant features must be dropped because they can lead to data leakage, where a feature reveals too much information about the target variable, effectively "giving away" the answer. In this study, the dropped features were nearly identical to the target variable, which is problematic as it skews the model's ability to properly assess the contribution of other features. Including such features inflates their importance while deflating the significance of other explanatory variables. This undermines the model's ability to capture meaningful relationships within the data, which is critical for understanding the true drivers of the outcome.

Additionally, irrelevant features were dropped because they add noise to the model, rather than valuable information. Such features do not contribute to predicting the target and instead make the model more complex, leading to poorer generalization to unseen data. Removing these irrelevant features allows the model to focus on the variables that matter, improving both interpretability and performance \cite{ge03}.

Imputing missing values with the mean or mode is particularly justified for several reasons. First, the preservation of data size is crucial for the reliability of the analysis. By imputing missing values with the mean for continuous variables or the mode for categorical variables, the study retains all available cases. This is vital to ensuring that the sample size remains robust, as removing rows with missing data could significantly limit the number of patients available for analysis, potentially undermining the study's conclusions \cite{sg02}.

In addition, imputation helps reduce bias that can arise from missing data. In the context of this dataset, missing values could be related to specific demographic or clinical characteristics that are essential for understanding the nuances of breast cancer treatment and outcomes. Simply dropping rows with missing values could lead to systematic differences between the included and excluded cases, ultimately skewing the results. By imputing missing values, this study provides a more accurate representation of the dataset, enabling the model to reflect the true relationships between treatment variables and patient outcomes without the confounding effects of missingness \cite{dvsm06,dp13}. 

Imputation methods that utilize the mean and mode specifically help maintain the overall distribution of the dataset, especially considering that the proportion of missing values in clinical datasets can be relatively small. Using mean imputation helps preserve the data's normal distribution, which is essential for meeting the assumptions of linear models employed in this study. This approach ensures that the imputed values do not introduce biases that could distort the analysis of treatment effects and patient outcomes.

The choice of encoding techniques for nominal and ordinal variables in this study is guided by the nature of the data and the requirements of linear modeling. Linear models, by design, assume a linear relationship between the input features and the target variable. This necessitates the transformation of categorical variables into a numerical format that aligns with these assumptions \cite{h01}.

One-hot encoding was selected for nominal features because these variables represent categories without any intrinsic order (e.g., race and ethnicity, treatment types). In a linear model, treating nominal variables as ordinal by assigning numeric values could lead the model to incorrectly infer a ranking or linear relationship where none exists. One-hot encoding resolves this issue by creating binary indicators for each category, allowing the linear model to treat each category independently without assuming any ordering.

For ordinal features, ordinal encoding was used. Ordinal variables possess a natural order (e.g., tumor grades, staging), and ordinal encoding allows the linear model to capture this rank-order relationship. By mapping these categories to integers, the encoding preserves the underlying structure of the data, which is appropriate for the linear model's assumptions. This ensures that the model can interpret the increasing or decreasing trend in the variable's effect on the outcome. Additionally, ordinal encoding reflects the fact that differences between levels may carry meaningful information for the linear model to exploit.
% The summary below shows the variables used in our analysis and categorizes them into their appropriate column types.

% \textbf{Dropped Columns}
% \begin{itemize}
%     \item Patient ID
%     \item Date of Birth (Days)
% \end{itemize}

% \textbf{Numeric Columns}
% \begin{itemize}
%     \item Recurrence event(s)
%     \item Days to last distant recurrence free assessment (from the date of diagnosis)
%     \item Days to Surgery (from the date of diagnosis)
% \end{itemize}

% \textbf{Nominal Columns}
% \begin{itemize}
%     \item Race and Ethnicity
%     \item ER
%     \item PR
%     \item HER2
%     \item Mol Subtype
%     \item Multicentric/Multifocal
%     \item Contralateral Breast Involvement
%     \item Lymphadenopathy or Suspicious Nodes
%     \item Skin/Nipple Involvement
%     \item Pec/Chest Involvement
%     \item Known Ovarian Status
% \end{itemize}

% \textbf{Ordinal Columns}
% \begin{itemize}
%     \item Menopause (at diagnosis)
%     \item Staging(Tumor Size) \# [T]
%     \item Staging(Nodes) \# (Nx replaced by -1) [N]
%     \item Staging(Metastasis) \# (Mx -replaced by -1) [M]
%     \item Tumor Grade(T) (Tubule)
%     \item Tumor Grade(N) (Nuclear)
%     \item Tumor Grade(M) (Mitotic)
% \end{itemize}

\subsection{Feature Selection}

Hyperparameter tuning was conducted using 5-fold cross-validation to find the configuration that resulted in the lowest root mean square error (RMSE). Cross-validation is a well-established method for model selection, allowing for more reliable estimation of model performance compared to a simple train-test split (Kohavi, 2001). After that, we introduced five random Gaussian variables and ran Lasso regression (Tibshirani, 1996) five times with different random states, using the optimal alpha parameter (Appendix A). By looking at the average importance of features across these runs, we aimed to identify the most significant variables while minimizing the variance of randomness in the Lasso model.  We visualized the top 20 features to identify features with importance significantly higher than the most important random variable, which is Random\_5 in this case.  We excluded all variables from the bottom of the ranking up until Race\_and\_Ethnicity\_White since the importance seems to be similar to Random\_5, which is contributing to the model just by chance.

\begin{figure*}[h]
\centering
\includegraphics[width=1\linewidth]{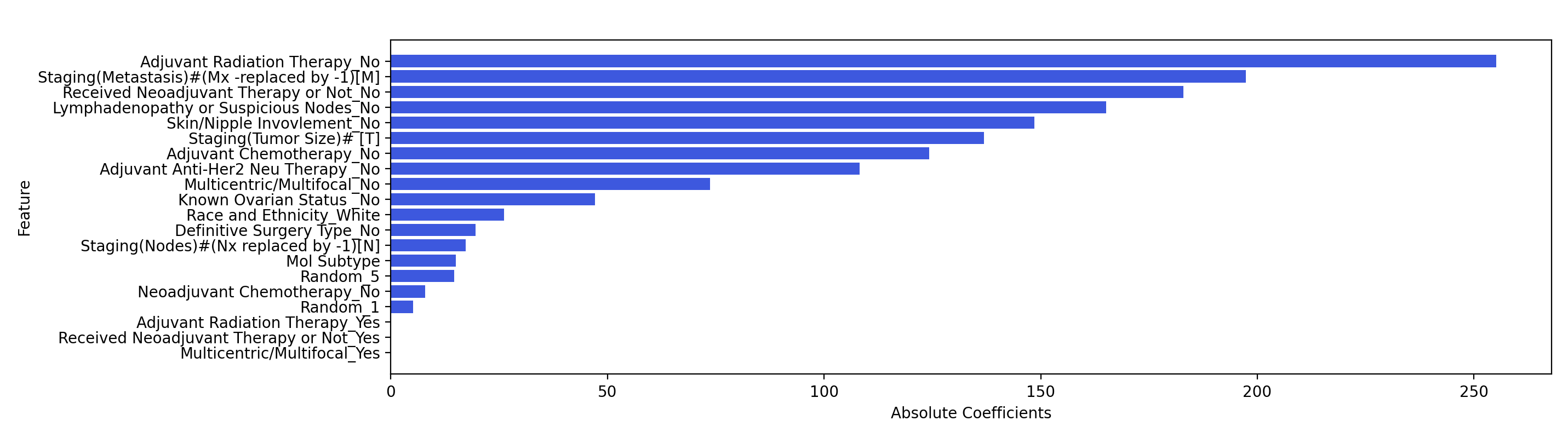}
\caption{Feature importance plot showing the absolute coefficients of various clinical and treatment variables in predicting breast cancer outcomes. The variables are ranked by their coefficient magnitude, with `Adjuvant Radiation Therapy\_No' having the strongest predictive power and `Multicentric/Multifocal\_Yes' having the least impact.}
\end{figure*}

\subsection{Causal Inference}
Traditional conditional independence methods rely on the assumptions of acyclicity (no feedback loop in causal graph) and absence of hidden common causes to isolate causal directions to the following three scenarios \cite{sgs01}.

\begin{figure}[h]
\centering
\includegraphics[width=1\linewidth]{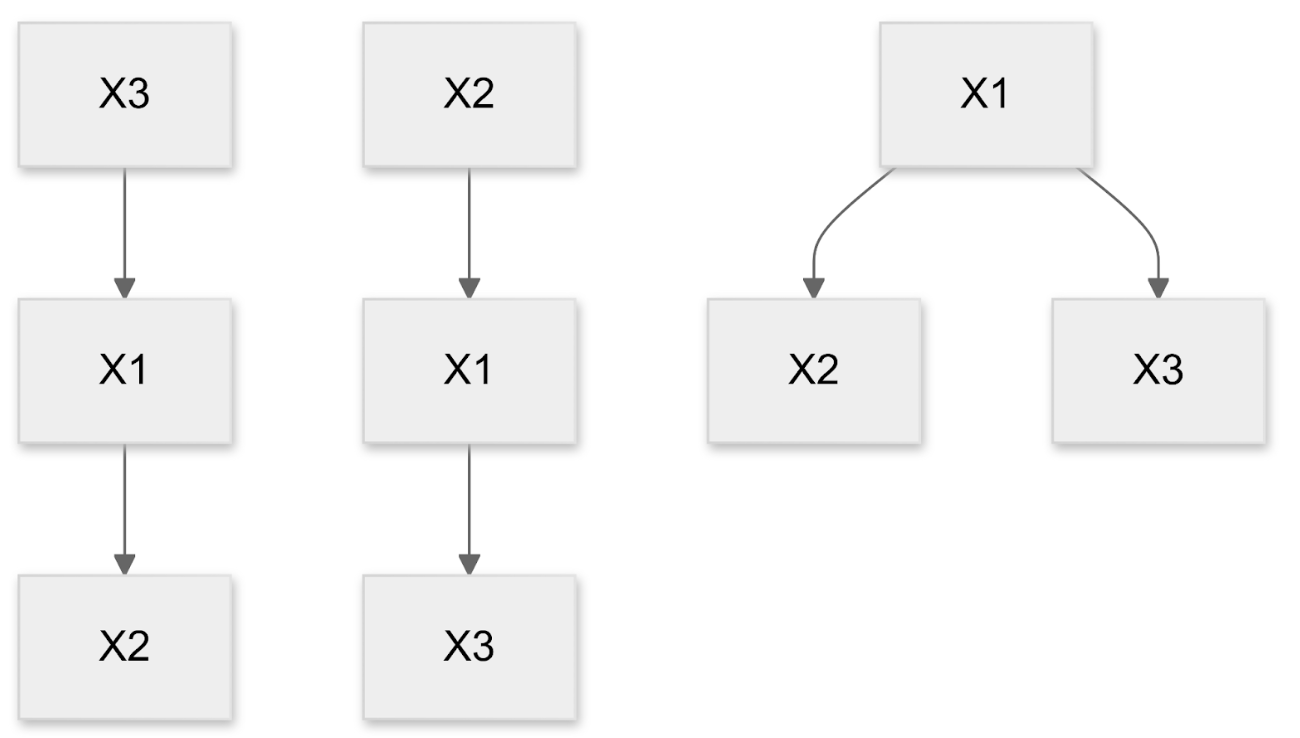}
\caption{Illustration of three possible causal structures under acyclicity and absence of hidden common causes assumptions}
\label{fig:yourlabel}
\end{figure}

However, we still do not know which of the three cases is correct, so the logical next step is to add further assumptions, non-Gaussian distribution and linearity, to exploit the asymmetry between regressing y against x and x against y, thereby identifying the correct causal scenario \cite{shhk06}.  

We utilized DirectLiNGAM \cite{sishkwhb11}, which operates on all of these assumptions combined.
DirectLiNGAM is an extension of the earlier proposed LiNGAM \cite{shhk06} model, which was one of the first to utilize linear non-Gaussian models for causal discovery. LiNGAM was designed to identify causal relationships by leveraging the non-Gaussian properties of the data. DirectLiNGAM builds on this approach by offering a more direct and efficient method for determining causal directions, simplifying the process and eliminating some of the iterative steps required in earlier models \cite{sishkwhb11}.

Given the underlying model below (non-Gaussianity and linearity assumptions met), we can see an asymmetry in regression outcomes based on the direction \cite{shhk06}.

\begin{gather*}
x_1 = e_1 \\
x_2 = b_{21}x_1 + e_2 \quad (b_{21} \neq 0) \\
e_1, \, e_2 \quad \text{non-Gaussian}
\end{gather*}

While regressing in the correct causal direction yields independence between residual and explanatory variable, when regressed in direction opposite of true causal direction, the explanatory variable and error term become dependent (both have e2). However, they are uncorrelated, and this is why distributions need to be non-Gaussian since uncorrelated Gaussian distributions suggest independence (both directions, same result).

\begin{gather*}
r_2 = x_2 - \frac{\text{cov}(x_2, x_1)}{\text{var}(x_1)}x_1 = x_2 - b_{21}x_1 = e_2 \\
x_1 (= e_1) \text { and } r_2 \text { are independent.}
\end{gather*}

\begin{gather*}
r_1 = x_1 - \frac{\text{cov}(x_1, x_2)}{\text{var}(x_2)}x_2 \\
= \left(1 - \frac{b_{21}\text{cov}(x_1, x_2)}{\text{var}(x_2)}\right)e_1 - \frac{b_{21}\text{var}(x_1)}{\text{var}(x_2)}e_2 \\
x_2 (= b_{21}e_1 + e_2) \text{ and } r_1 \text{ are dependent,} \\
\text{although they are uncorrelated. }
\end{gather*}

With these assumptions, DirectLiNGAM works by performing pairwise regression to identify variables in a topological order, starting with the variable at the top, denoted as \( x_1 \). After identifying \( x_1 \), the method removes its effect on the subsequent variables by computing residuals, which involves regressing \( x_2 \) on \( x_1 \) and \( x_3 \) on \( x_1 \). The process continues by regressing the remaining residuals on each other to uncover the causal structure among the variables. Once this structure is established, the adaptive Lasso is employed to eliminate weak connections, refining the model and enhancing the robustness of the causal relationships identified \cite{sishkwhb11}.

\begin{gather*}
x_i = \sum_{k(j) < k(i)} b_{ij} x_j + e_i
\end{gather*}

Below is a summary of the assumptions of DirectLiNGAM:
Acyclicity
Linearity
Non-Gaussian
Absence of hidden common causes

We ensure each assumption is met in this study one by one. \\

\textbf{Acyclicity} \\
The variables in the dataset follow a natural temporal order. Patient demographics, such as age and gender, precede the onset of medical conditions, which are followed by treatments. This ensures that causality flows forward—conditions lead to treatments, not the reverse—eliminating the possibility of cyclical feedback where treatments could influence pre-existing conditions. \\

\textbf{Non-Gaussianity} \\
We used the Shapiro-Wilk test, where the null hypothesis of variable error is Gaussian, was rejected for all features (p-value was 0.00). \\

\textbf{Linearity} \\
We already know from Lasso regression that the outcome variable is a linear combination of the extracted features. Another point of interest is the nature of relationships between the features themselves since LiNGAM successively performs pairwise regression among features (or residuals from previous steps).  As seen in the correlation matrix below, we can observe some linear relationships among features.

\begin{figure*}[h]
\centering
\includegraphics[width=\textwidth]{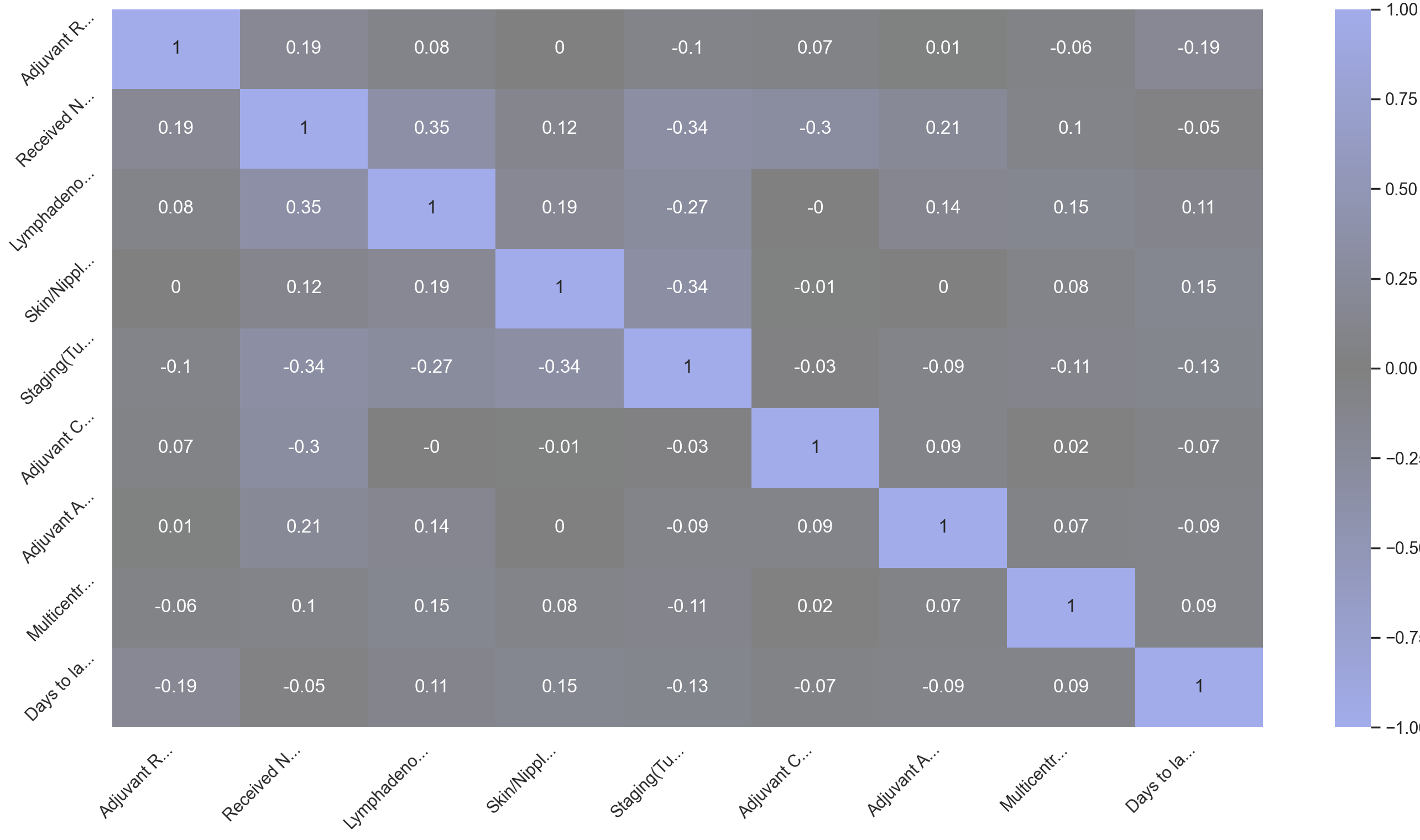}
\caption{Correlation matrix showing the relationships between different clinical variables in a medical study. The values range from -1 to 1, with blue colors indicating positive correlations and darker colors indicating negative or weaker correlations. The diagonal shows perfect correlation (1.0) as variables correlate perfectly with themselves.}
\label{fig:yourlabel}
\end{figure*}

\textbf{Absence of hidden common causes} \\
In order to test the absence of hidden common causes, we applied the error independence test where the null hypothesis is that residuals and explanatory variables are independent in each pairwise regression.

\begin{figure*}[h]
\centering
\includegraphics[width=1\linewidth]{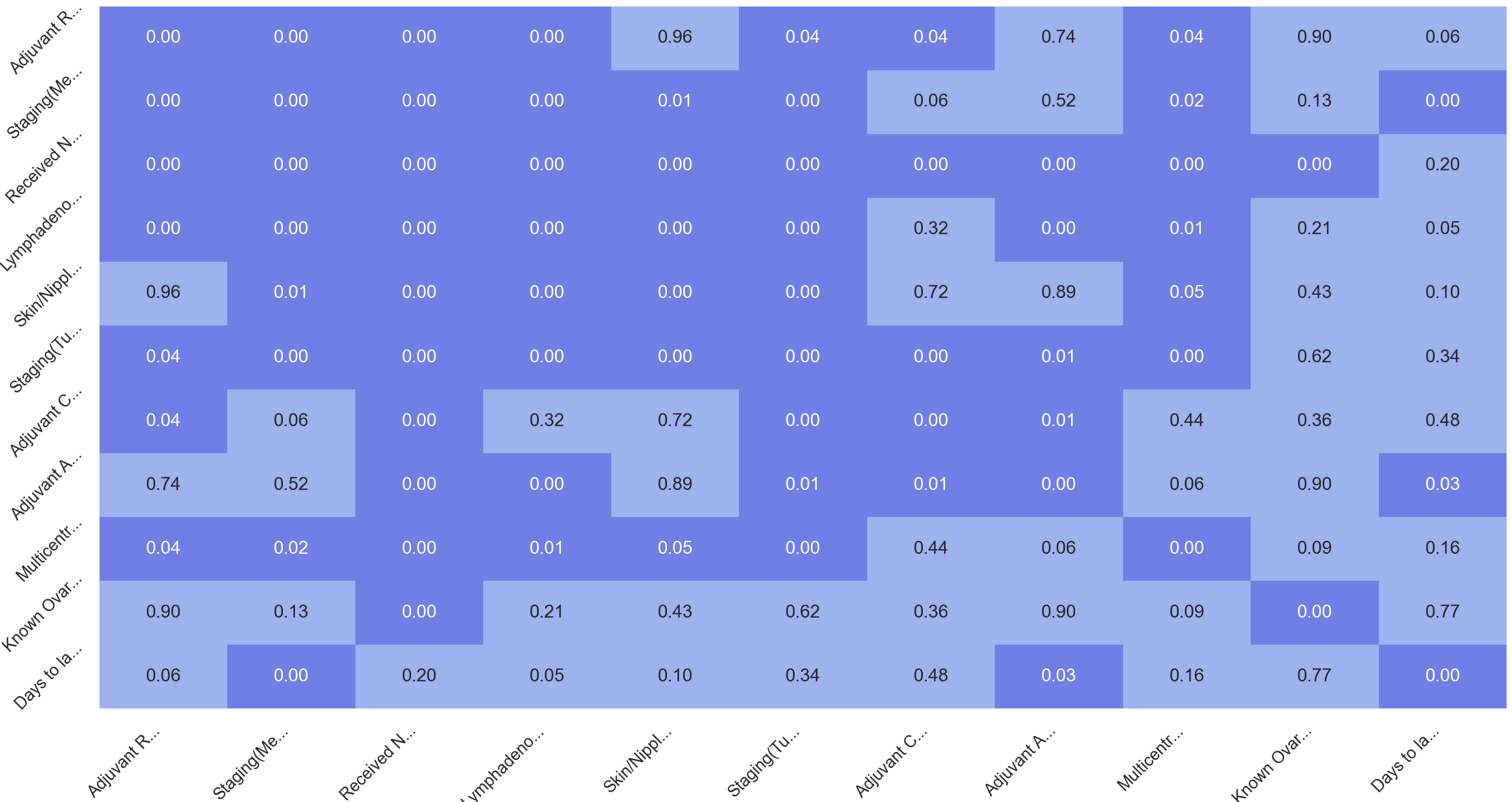}
\caption{P-value matrix displaying the statistical significance of relationships between medical variables. The values range from 0 to 1, with lighter colors indicating more statistically significant relationships (lower p-values). Values closer to 0.05 or less typically indicate statistical significance.}
\label{fig:yourlabel}
\end{figure*}

Although many features were included in the analysis, as suggested by the occurrence of p-values of 0.00, this assumption is violated in many pairs.  Therefore,  we used the LiNGAM RCD, which has the same set of assumptions as DirectLiNGAM, except it allows for hidden common causes. It outputs “a causal graph where a bi-directed arrow indicates the pair of variables that have the same latent confounders, and a directed arrow indicates the causal direction of a pair of variables that are not affected by the same latent confounder” (Maeda \& Shimizu, 2020).

\subsection{Evaluation}

To systematically evaluate the robustness of our estimates, we apply the backdoor criterion and test them against three refutation techniques: adding a random common cause, using a subset of data, and employing a placebo treatment.

We opted to evaluate the causal effects using the backdoor criterion and refutation techniques for several reasons. One key advantage of using the backdoor criterion is that it provides a structured, well-established framework for isolating causal relationships by explicitly accounting for confounders. This approach allows us to verify the robustness of our findings through an independent and complementary method, rather than relying solely on LiNGAM's assumptions.

Javidian and Valtorta (2018) provided a concise summary of the backdoor-criterion concept originally introduced by Hitchcock and Pearl (2001). They explain that the backdoor-criterion is satisfied by "a set of variables Z, relative to an ordered pair of variables (\( X_i \), \( X_j \)) in a Directed Acyclic Graph (DAG) G if:
(i) no node in Z is a descendant of \( X_i \)
(ii) (ii) \( Z \) blocks every path between \( X_i \) and \( X_j \) that contains an arrow pointing into \( X_i \). In the graph below, \( S_1 = \{X_3, X_4\} \) and \( S_2 = \{X_4, X_5\} \) would qualify under the back-door criterion, but \( S_3 = \{X_4\} \) would not because \( X_4 \) does not d-separate \( X_i \) from \( X_j \) along the path \( (X_i, X_3, X_1, X_4, X_2, X_5, X_j) \).

\begin{figure}[h]
\centering
\includegraphics[width=1\linewidth]{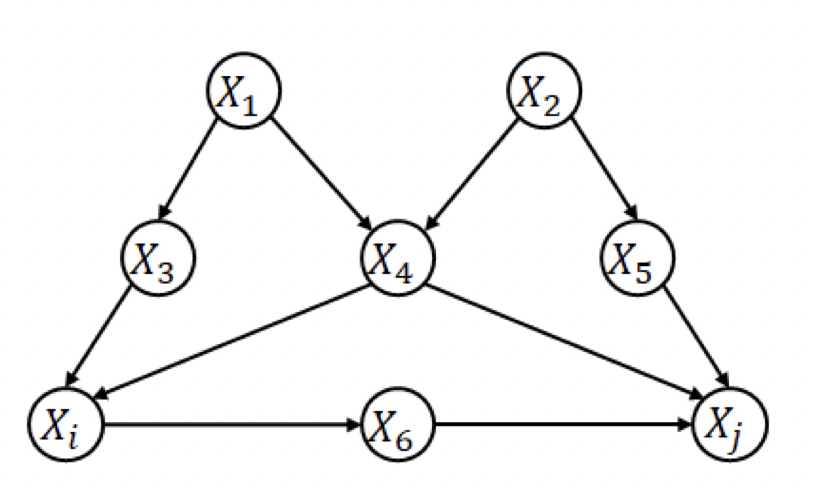}
\caption{Backdoor criterion example}
\label{fig:yourlabel}
\end{figure}

By applying the backdoor criterion, we can effectively isolate the causal effect of \( X_i \) on \( X_j \). This is achieved by ensuring that any confounding influences are accounted for through the selected variables in Z. Consequently, we reduce the risk of spurious associations that could arise from unblocked pathways, allowing us to obtain a clearer understanding of the direct causal relationship between the treatment and the outcome. This isolation is crucial for accurately interpreting the effects of our variables of interest, thereby strengthening the validity of our findings.

As for the refutation techniques, we have the following null hypotheses: adding a random common cause between the treatment and sink variables or using a subset of data should not yield a new estimated effect that is too different, and using a placebo treatment instead should bring down the new estimate to near zero.  The new estimates are compared against the original ones, and a p-value is returned for each null hypothesis.

By introducing a random common cause, we can assess whether the original effect remains consistent in the presence of additional noise. Using a subset of data allows us to examine the generalizability of the findings across different sample sizes and conditions. This method is grounded in propensity score matching and other techniques for controlling confounders as discussed by Rosenbaum and Rubin (1983), and is also part of the refutation suite in Sharma and Kiciman (2020), where it is used to evaluate whether causal effects are sensitive to data sampling variations. Lastly, the placebo treatment serves as a critical control, helping to determine whether the observed effects are indeed causal or merely artifacts of the analysis. Rosenbaum (2002) emphasizes the importance of placebo treatments in observational studies to ensure that no spurious correlations are driving the observed results. Collectively, these techniques enhance our confidence in the causal relationships identified, ensuring that they are not sensitive to specific assumptions or data conditions.

With the causal structure from LiNGAM, we estimated the causal effect and applied the refutation techniques for each variable of interest respectively using DoWhy, “an open-source Python library” for causal inference provided by Sharma and Kiciman (2020).

\section{Results}
Our analysis identified Adjuvant Anti-Her2 Neu Therapy and Skin/Nipple involvement as significant factors influencing local recurrence-free survival. Specifically, Adjuvant Anti-Her2 Neu Therapy was found to increase the number of days to the last local recurrence-free assessment by 169 days, on average. In contrast, Skin/Nipple involvement decreased this duration by 351 days, on average.
Below is a graph indicating the causal structure of the dataset, with directed and bi-directed arrows representing causal effect and existence of hidden common causes, respectively.  

\begin{figure*}[h]
\centering
\includegraphics[width=\textwidth]{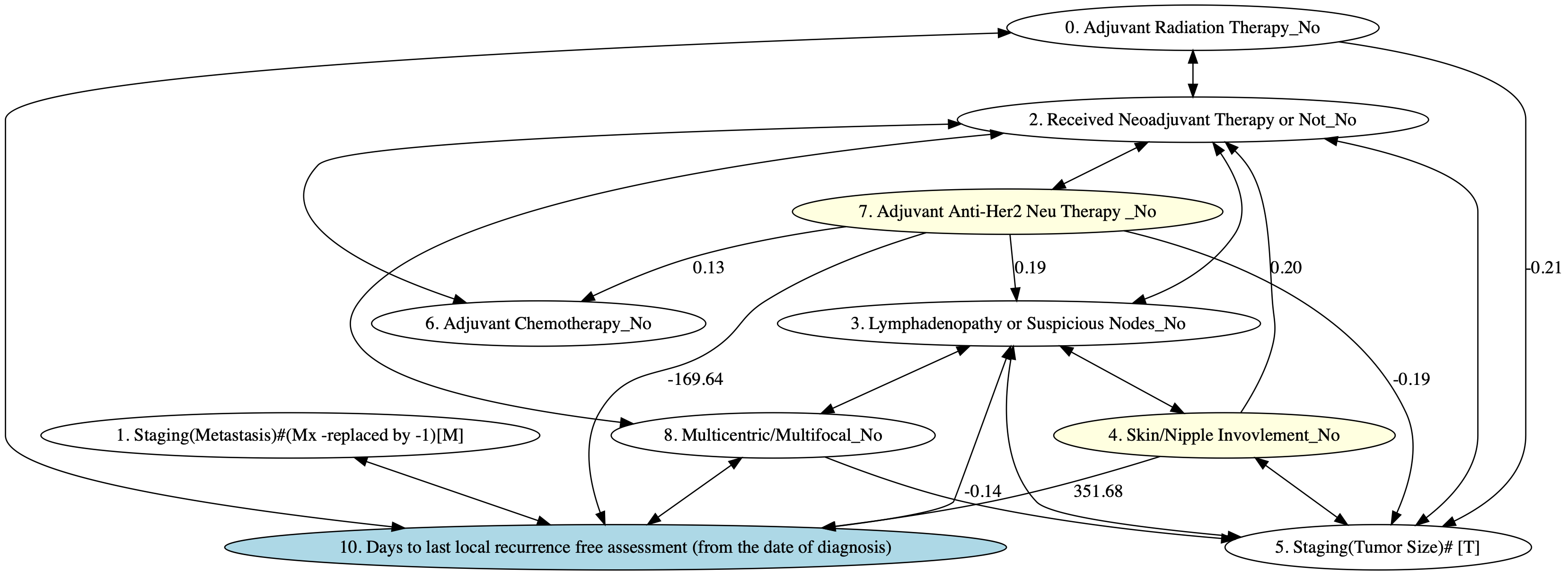}
\caption{LiNGAM (Linear Non-Gaussian Acyclic Model) causal graph depicting the structural relationships between breast cancer variables. The directed edges represent causal relationships between variables, with edge weights indicating the strength and direction of causal effects. The node numbered 10 (Days to last local recurrence free assessment) appears to be a key outcome variable. LiNGAM specifically assumes non-Gaussian distributions and uses this to determine causal direction between variables.}
\label{fig:yourlabel}
\end{figure*}

The causal effects are summarized in the table below.

\begin{table*}[ht]
\centering
\resizebox{\textwidth}{!}{
\begin{tabular}{|l|l|l|}
\hline
\textbf{From} & \textbf{To} & \textbf{Effect} \\ \hline
Skin/Nipple Involvement\_No & Received Neoadjuvant Therapy or Not\_No & 0.204 \\ \hline
Adjuvant Anti-Her2 Neu Therapy\_No & Lymphadenopathy or Suspicious Nodes\_No & 0.1862 \\ \hline
Adjuvant Radiation Therapy\_No & Staging(Tumor Size)\# [T] & -0.2108 \\ \hline
Adjuvant Anti-Her2 Neu Therapy\_No & Staging(Tumor Size)\# [T] & -0.1886 \\ \hline
Multicentric/Multifocal\_No & Staging(Tumor Size)\# [T] & -0.144 \\ \hline
Days to last local recurrence free assessment (from the date of diagnosis) & Staging(Tumor Size)\# [T] & -0.0002 \\ \hline
Adjuvant Anti-Her2 Neu Therapy\_No & Adjuvant Chemotherapy\_No & 0.1254 \\ \hline
Skin/Nipple Involvement\_No & Days to last local recurrence free assessment (from the date of diagnosis) & 351.6813 \\ \hline
Adjuvant Anti-Her2 Neu Therapy\_No & Days to last local recurrence free assessment (from the date of diagnosis) & -169.638 \\ \hline
\end{tabular}
}
\caption{Summary of causal effects identified through LiNGAM analysis in breast cancer treatment data. The table shows direct causal relationships between clinical variables and their effect sizes. Notable findings include the strong positive effect of no skin/nipple involvement on recurrence-free assessment time (351.68 days), and the substantial negative impact of no adjuvant Anti-Her2 Neu therapy on recurrence-free assessment time (-169.64 days). The 'From' column indicates the causal variable, the 'To' column shows the affected variable, and the 'Effect' column quantifies the strength and direction of each causal relationship.}
\end{table*}

As highlighted in the causal figure above, the main contributors to local recurrence free assessment are Adjuvant Anti-Her2 Neu Therapy and Skin/Nipple Involvement.

We subsequently tested our results with the backdoor criterion and the three refutation techniques. Below are the results of the evaluation. \\

\begin{table*}[ht]
\centering
\resizebox{\textwidth}{!}{
\begin{tabular}{|l|l|l|l|l|}
\hline
\textbf{Variable} & \textbf{Refutation Technique} & \textbf{Estimated Effect} & \textbf{New Effect} & \textbf{p-value} \\ \hline
Adjuvant Anti-Her2 Neu Therapy & Add a random common cause & -168.54 & -168.39 & 0.92 \\ \hline
Adjuvant Anti-Her2 Neu Therapy & Use a subset of data & -168.54 & -165.34 & 0.96 \\ \hline
Adjuvant Anti-Her2 Neu Therapy & Use a Placebo Treatment & -168.54 & -11.65 & 0.9 \\ \hline
Skin/Nipple Involvement & Add a random common cause & 350.82 & 351.05 & 0.88 \\ \hline
Skin/Nipple Involvement & Use a subset of data & 350.82 & 349.71 & 0.9 \\ \hline
Skin/Nipple Involvement & Use a Placebo Treatment & 350.82 & 1.75 & 0.9 \\ \hline
\end{tabular}
}
\caption{Sensitivity analysis of key causal relationships identified in the breast cancer treatment study. The table compares original causal effect estimates with results from three different refutation techniques: adding random common causes, using data subsets, and applying placebo treatments. For both Adjuvant Anti-Her2 Neu Therapy and Skin/Nipple Involvement, the estimated effects remain largely stable under the first two refutation methods (as indicated by high p-values), but show substantial changes under the placebo treatment condition. This suggests that while the causal relationships are robust to common confounders and sampling variations, they may be sensitive to treatment-specific effects.}
\end{table*}

\textbf{1. Adjuvant Anti-Her2 Neu Therapy}  
\begin{itemize}
    \item Adding a Random Common Cause: The estimated effect was initially -168.54. After adding a random common cause, the new effect remained almost unchanged at -168.39, with a p-value of 0.92, suggesting that the initial estimate is robust against random confounding. 
    \item Using a Subset of Data: The estimated effect slightly changed to -165.34 with a p-value of 0.96, indicating that the estimate is consistent when using different subsets of the data. 
    \item Using a Placebo Treatment: This refutation resulted in a much smaller effect size of -11.65 with a p-value of 0.9, implying that the causal effect is likely not due to random chance. 
\end{itemize}

\textbf{2. Skin/Nipple Involvement} 
\begin{itemize}
    \item Adding a Random Common Cause: The estimated effect was 350.82, and after adding a random common cause, it remained nearly the same at 351.05, with a p-value of 0.88. This consistency suggests that the causal effect is not significantly affected by potential confounders.
    \item Using a Subset of Data: The effect size changed slightly to 349.71 with a p-value of 0.9, demonstrating stability across different data subsets.
    \item Using a Placebo Treatment: The new effect dropped to 1.75 with a p-value of 0.9, similar to the findings with the Adjuvant Anti-Her2 Neu Therapy, implying that the causal effect is likely not due to random chance.
\end{itemize}

The findings for Adjuvant Anti-Her2 Neu Therapy show that the estimated effect remained stable when accounting for a random common cause and using different data subsets, indicating robustness against confounding and consistency across data. For Skin/Nipple Involvement, the estimated effect also remained consistent when a random common cause was introduced and across data subsets, suggesting the causal effect is not significantly impacted by potential confounders. In both cases, the placebo treatment significantly reduced the effect sizes, further implying that the observed causal effects are likely not due to random chance.

\section{Discussion}
These findings highlight the beneficial effects of Adjuvant Anti-Her2 Neu Therapy in extending local recurrence-free survival, while also pointing to the severe negative impact of Skin/Nipple involvement. The positive impact of Adjuvant Anti-Her2 Neu Therapy observed in this study underscores its critical role in extending recurrence-free survival in breast cancer patients, particularly those with Her2-positive tumors \cite{hpco19}. Conversely, the negative association between Skin/Nipple involvement and recurrence-free survival points to a higher risk of early recurrence in patients presenting with this condition. This suggests that Skin/Nipple involvement serves as a key prognostic marker, indicating the need for more aggressive or targeted therapeutic strategies. Patients with this feature may benefit from intensified treatment regimens or closer post-treatment monitoring to mitigate the elevated risk of recurrence \cite{dmtbh15}.

Beyond the impact these findings have on the treatment planning and prognosis of breast cancer patients, they also create opportunities for further research. In particular, there is an urgent need to optimize interventions for high-risk groups, such as those with Skin/Nipple involvement, as currently "nipple-sparing mastectomy (NSM) is an increasingly common procedure [for this purpose, but] concerns exist regarding its oncological safety due to the potential for residual breast tissue to harbor occult malignancy or future cancer" \cite{dmtbh15}. Additionally, the methodology used in this study could be applied to other observational patient datasets in the field, offering a broader scope for discovering causal relationships and improving patient outcomes. \cite{iaahr22} lists out 20 breast cancer observational datasets, which can easily be studied with our methods:

%\begin{table}[ht]
%\centering
%\resizebox{250pt}{!}{
%\begin{tabular}{|l|l|}
%\hline
%\textbf{Dataset Name} & \textbf{Link} \\ \hline
%Cancer Waiting Times & \href{https://data.world/datasets/breast-cancer}{data.world/datasets/breast-cancer} \\ \hline
%NKI Breast Cancer Data & \href{https://data.world/datasets/breast-cancer}{data.world/datasets/breast-cancer} \\ \hline
%ISPY1 Trial & \href{https://data.world/datasets/breast-cancer}{data.world/datasets/breast-cancer} \\ \hline
%Mammographic Masses & \href{https://data.world/datasets/breast-cancer}{data.world/datasets/breast-cancer} \\ \hline
%UTA4 Datasets & \href{https://data.world/datasets/breast-cancer}{data.world/datasets/breast-cancer} \\ \hline
%Wisconsin Diagnostic Dataset & \href{https://www.kaggle.com/uciml/breast-cancer-wisconsin-data}{kaggle.com/uciml/breast-cancer-wisconsin-data} \\ \hline
%UCI Breast Cancer & \href{https://data.world/uci/breast-cancer}{data.world/uci/breast-cancer} \\ \hline
%SEER Breast Cancer Data & \href{https://ieee-dataport.org/open-access/seer-breast-cancer-data}{ieee-dataport.org/seer-breast-cancer-data} \\ \hline
%BreakHis Database & \href{https://web.inf.ufpr.br/vri/databases/breast-cancer-histopathological-database-breakhis/}{web.inf.ufpr.br/vri/databases/breakhis} \\ \hline
%ATP5B Expression Data & \href{https://www.frontiersin.org/articles/10.3389/fgene.2021.652474/full}{frontiersin.org/articles/10.3389/fgene.2021.652474} \\ \hline
%Multi-Modal Dataset & \href{https://data.world/datasets/breast-cancer}{data.world/datasets/breast-cancer} \\ \hline
%Wisconsin Original Dataset & \href{https://archive.ics.uci.edu/ml/datasets/breast+cancer+wisconsin}{archive.ics.uci.edu/ml/datasets/breast+cancer+wisconsin} \\ \hline
%Scikit-learn Dataset & \href{https://scikit-learn.org/stable/modules/generated/sklearn.datasets.load_breast_cancer.html}{scikit-learn.org/datasets/breast-cancer} \\ \hline
%CBIS-DDSM Images & \href{https://www.kaggle.com/awsaf49/cbis-ddsm-breast-cancer-image-dataset}{kaggle.com/cbis-ddsm-breast-cancer} \\ \hline
%USF Mammography DB & \href{http://marathon.csee.usf.edu/Mammography/Database.html}{marathon.csee.usf.edu/Mammography} \\ \hline
%GDC Cancer Data & \href{https://portal.gdc.cancer.gov}{portal.gdc.cancer.gov} \\ \hline
%cBioPortal & \href{https://www.cbioportal.org/}{cbioportal.org} \\ \hline
%\end{tabular}
%}
%\caption{List of Breast Cancer Datasets with Access Links}
%\end{table}

\section{Conclusion}
In conclusion, this study demonstrates the value of applying causal inference methods to observational data in breast cancer research. Our analysis of the Duke MRI Breast Cancer dataset, using the Linear Non-Gaussian Acyclic Model (LiNGAM), revealed significant causal relationships affecting local recurrence-free survival. Specifically, Adjuvant Anti-Her2 Neu Therapy was found to increase the number of days to the last local recurrence-free assessment by an average of 169 days, while Skin/Nipple involvement decreased this duration by an average of 351 days.

%\section{Appendix}

%\appendix
%\section{Additional Figures}
%\begin{figure}[h]
%\centering
%\includegraphics[width=1\linewidth]{Hyperparameter Tuning.png}
%\caption{Optimal alpha value in lasso regression model}
%\label{fig:appendix_figure}
%\end{figure}

%%%%%%%%% REFERENCES

\bibliographystyle{plain}
\bibliography{bibtex}

\end{document}